# Auto-Regressive Model with Exogenous Input (ARX) Based Traffic Flow Prediction

**JUN YING.,[1] XIN DONG.,[2] BOWEI LI.,[3] and ZIHAN TIAN.[4]**

Department of Civil and Environmental Engineering, University of Michigan, Ann Arbor, MI, 48105; e-mail: [1] yingjun@umich.edu [2] xindong@umich.edu [3] jacklbw@umich.edu [4] zti@umich.edu

**ABSTRACT**

Traffic flow prediction is widely used in travel decision making, traffic control, roadway system planning, business sectors, and government agencies. ARX models have proved to be highly effective and versatile. In this research, we investigated the applications of ARX models in prediction for real traffic flow in New York City. The ARX models were constructed by linear/polynomial or neural networks. Comparative studies were carried out based on the results by the efficiency, accuracy, and training computational demand of the algorithms.

**INTRODUCTION**

Traffic flow prediction is significant and challenging for route planning and transportation infrastructure planning (Zhang et al. [2008]). The Federal Highway Administration (FHWA) developed a real-time traffic prediction system for route planning and transportation management (Zhang et al.20 [2008]). New York Metropolitan Transportation Council (NYMTC) collected traffic volume data every year and applied it to traffic demand forecasting and transportation infrastructure planning (NYMTC [2018]). Liebig et al. [2017] developed a route planning system for individual travelers utilizing future traffic conditions and applied it to a real-world scene in Dublin, Ireland. The traffic flows are generally a location related dynamic time series that depends on external factors, such as weather, road conditions, or internal factors as past traffic flows. ARX model, which involves both external and internal factors, can potentially be a suitable tool for traffic flow prediction. The general form of ARX model is Eq. (1) (Billings et al. [1989], Haviluddin and Dengen [2016], Menezes Jr and Barreto [2008]).

$$u(t) = G(p(t), p(t-1), \ldots, p(t-t_{d,p}), u(t-1), u(t-t_{d,u})) \quad (1)$$

Where $u(t)$ and $p(t)$ are output and input at time t, respectively. $t_{d,p}$ and $t_{d,u}$ are the maximum time delays corresponding to output and input variables, respectively. $G(\cdot)$ is a function that represents the ARX model. The core problem in this research is to find the appropriate form and determine function $G(\cdot)$. The solution can be non-unique.

In this research, the application of ARX model will be investigated, potential exogenous inputs include the local weather and neighbor traffic information.

## RELATED WORKS

Traffic data prediction is a time series prediction that can be dealt with in general auto-regressive models, e.g. (vector) auto-regressive model ((V)AR) (Chandra and Al-Deek [2009], Abadi et al. [2014]), auto-regressive integrated moving average model (ARIMA) (Dehuai Zeng et al. [2008], Groschwitz and Polyzos [1994], auto-regressive integrated moving average errors model (ARIMAX) (Williams [2001]), stacked auto-encoder (Lv et al. [2014]), Hybrid models (Dehuai Zeng et al. [2008]). A limited time ahead prediction was successfully implemented in the research. However, exogenous inputs that represent the influence of external factors are generally not considered in the previous research. Further, the previous research is mostly about limited steps ahead prediction. This research investigates the effect of potential exogenous inputs and the feasibility of free run prediction.

## METHOD AND APPROACH

In this research, we firstly implement data clustering for the data set. By clustering analysis, we expect a clear picture on the behaviors of the time series in each cluster. Then we carry out ARX training, as shown above; the key is to find the function $G(\cdot)$. Generally, the work is carried out by firstly assuming the form of $G(\cdot)$, and then implementing machine learning algorithms to determine the rest unknowns in the form. Details are as follows.

### Data Clustering

K-Means clustering solves the problem of identifying subgroups in a group of unsupervised data. Usually, the task is to divide n observations into K clusters based on Euclidean Distance Metric, and a vector is predicted containing cluster indices of each observation. However, Euclidean Distance Metric can only capture spatial correlations among locations but cannot capture the pattern similarities of time series in multiple traffic locations. So, for time series data, we use Dynamic-Time-Warping (DTW) method to calculate the distance matrix and embed it in K-Means clustering to improve clustering accuracy. The distance calculated by DTW method is generally defined as Eq. (2).

$$\gamma(i,j) = d(q_i, c_j) + \min \{\gamma(i-1, j-1), \gamma(i-1, j)\gamma(i, j-1)\} \quad (2)$$

Where i, j are respectively two time series, cumulative distance between i and j is defined as $\gamma(i,j)$ and $d(q_i, c_j)$ is cummulative distance found at current cell ($q^{th}$ value of series i and $c^{th}$ value of series j).

### ARX Training

In general, the forms of ARX can be linear/polynomials (Billings et al. [1989]), wavelet (Billings and Wei [2005]), and neural networks (Diaconescu [2008]). Normally, the general forms can lead to ARX model with more generality, but less physical meaning. In practical applications, any available knowledge on the system behind the problem can be used to inform the form selection. By doing so the results usually have more physical meaning, but less generality. One may need to balance the two aspects so that the ARX model can achieve higher accuracy but without being trivial. In addition, parameters such as model order (maximum time delay, $t_{d,p}$ and $t_{d,U}$ in Eq. (1)), and maximum polynomial order for polynomial form, etc., can be directly selected, based on available knowledge. Normally, it is always preferable to start with simple models having a reasonable number of terms. Model determination typically includes model structure determination and coefficient calibration. The two parts can be carried out separately or simultaneously. For model structure determination, one may use forward selection, backward elimination, all subset regression, etc. (Efron et al. [2004]). For coefficient calibration, available methods include least square regression, maximum likelihood method, Bayesian method, back-propagation, etc. The solution can be either close formed or must be obtained by numerical search.

In this research, the most popular linear/polynomial and neural network ARX models were considered. For linear/polynomial ARX models, the LARs, which belongs to the forward selection algorithm, which will be implemented to determine the model structure. Then, OLS, RR, and Lasso (with/without cross-validation), will be adopted to further calibrate model parameters. For the neural network ARX model, SRNN and LSTM were implemented, which were trained by LM and Adam algorithms, respectively. The training procedures are shown in Figure 1.

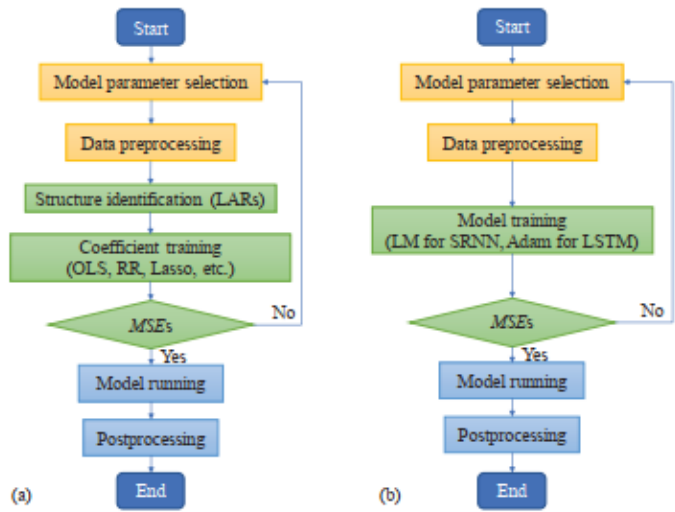

**Figure 1 : Flow-Chart for (a) Linear/Polynomial, and (b) RNN ARX**

**Feature Engineering**

By Eq. (1), the features can be previous outputs and inputs. Thus, the features for every current output u(t) includes previous output, $u(t-1)$, , $u(t-2)$, ..., $u(t-t_{d,u})$, and all possible inputs $p(t), p(t-1), ..., p(t-t_{d,p})$. The inputs include traffic volume data for neighbor roads, weather data (humidity, pressure, temperature, and weather descriptions).

**Evaluation**

The evaluation of the trained models includes by relative mean square error (MSE$_r$) metrics and training time. The MSE$_r$s includes MSE$r_a$, MSE$r_{tr}$, MSE$r_{te}$ and MSE$r_m$, which indicate MSE$_r$ for all, training, test, and free run predictions, and generally defined as Eq. (3).

$$MSE_r = \frac{||u_{ob}(t) - u_{pred}(t)||^2}{||u_{ob}(t)||^2} \qquad (3)$$

where $u_{ob}(t)$ and $u_{pred}(t)$ are observed and predicted outputs, respectively. The metric MSE$r_a$, gives an overview on how a specific model performs when learning the time series. While the difference between MSE$r_{tr}$, MSE$r_{te}$ indicates if overfitting happens. Further, training time is also included to evaluate models.

## EXPERIMENTS AND RESULTS
### Data Description

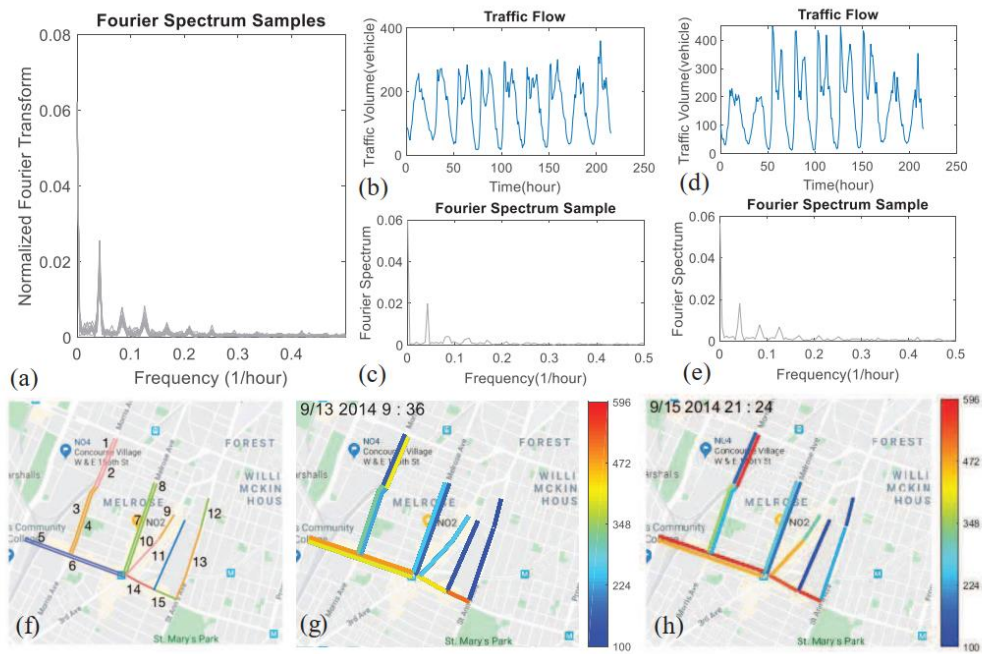

**Figure 2: Data visualization: (a) Fourier spectrum Samples, (b) traffic flow sample 1, (c) Fourier spectrum sample 1, (d) traffic flow sample 2, (e) Fourier spectrum sample 2. (f) Road distribution (g-h) Traffic volume heat map at two representative time points**

This data set is from NYC open data, traffic volume counts are collected by the Department of Transportation for New York to validate the New York Best Practice

Model. The range of the timestamp is from 09/13/2014 to 04/15/2018. 2272 road segments are involved in data collection, and each day is divided into 24 time periods, with a uniform sampling frequency of 1 hour.

Each segment has about 10-days records. To visualize the traffic volume information, we located 17 relatively concentrated segments. This volume data shows periodic patterns. The data is visualized and shown in Figure 2. Figure 2 (a) shows all Fourier spectrum samples of 17 segments volume. It can be seen from the samples that the data generally exhibits a clear pattern, while also with some randomness. Two representative traffic flow data features are highlighted, with the time series in Figure 2 (b)(d) and the corresponding Fourier spectra shown in Figure 2(c)(e), respectively. Figure 2 (d) is different from Figure 2 (b) in that less traffic volume occurs during weekends. The original road map and traffic volume heat map at two representative time points are shown in 2 (f-h). We here implement time series clustering for the data by using DTW based K-Means method, with the optimum K value determined by Calinski-Harabasz criterion. 7 clusters lead to the optimal performance. All the time series have double and single crest per cycle, which shows a clear weekdays-weekends pattern. Traffic flow is highly concentrated in commuting time during weekdays, which cannot be seen on weekends. Single crests during weekends does not have obvious morning and night peak hours. This observation is consistent with common sense. The traffic flow volume crests indicate road types of clusters. Based on 6$^{th}$ edition of Highway Capacity Manual (Manual [2000]), the adjusted saturation flow rate is about 2000 vehicle/lane/hour for urban or suburban street, and the adjusted saturation flow rate is about 1000 vehicle/lane/hour for rural streets. Some double crests per period have two significant different crest values, which shows a strong tidal pattern, i.e. the traffic flow reaches crest during morning peak hours from home to workplace, and night peak hours in opposite direction. One may further infer the buildings or communities' type around the road, e.g. workplaces, and residential area, etc. Based on the analysis, road type descriptions are listed in Table 1.

**Table 1: Feature Table of Clusters**

| Cluster Number | 1 | 2 | 3 | 4 | 5 | 6 | 7 |
|---|---|---|---|---|---|---|---|
| Crest Value | 1400 | 4000 | 700 | 1300 | 3000 | 4200 | 3500 |
| Road Type Label | S | U | R | S | U | U | U |
| Weekday-weekend Pattern | N | Y | N | Y | Y | N | Y |
| Majority Travels Type Label | 0 | 1 | 0 | 1 | 1 | 0 | 1 |
| Tidal Pattern | N | N | Y | Y | Y | N | Y |

Note: For road types, U is urban, S is suburban, R is rural; For pattern, Y is yes, N is no.
Majority Travels Type, 1 is Commuter, 0 is Otherwise

The hourly recorded weather data were obtained from (Kaggle [2019]). Data as pressure, humidity, and pressure that recorded quantitatively can be directly used. Weather descriptions that recorded in qualitative manner were converted to be a real number from [0; 1] by averaging independent experience-based evaluation by all group members. The corresponding independent evaluation converts weather descriptions to the degree of ease to drive through. The weather data is shown in Figure 3.

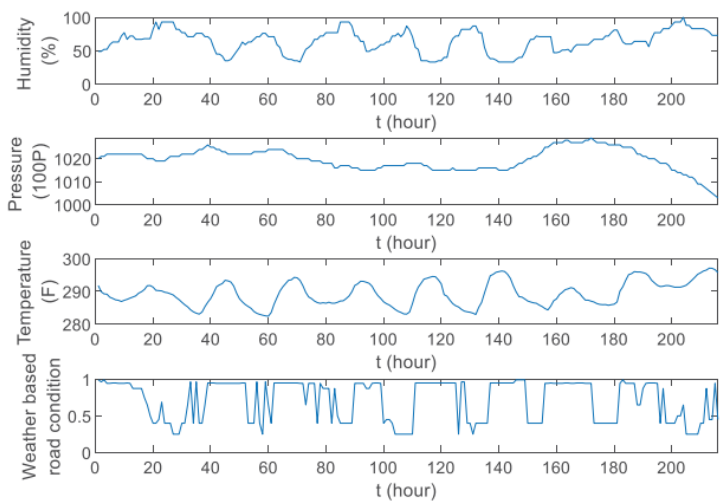

**Figure 3 Weather Data**

**Linear Auto-regressive Model**

In this section, we implemented time series learning by linear auto-regressive model. We applied LARs for structure learning and OLS, RR, and Lasso for weights learning. Maximum time delay td of 4 is considered for both inputs $t_{d,p}$ and outputs $t_{d,u}$. The linear auto-regressive model by OLS without exogenous input and validation case (AR OLS) is selected as the baseline case. For cases with validation, training, validation, and test sets count 50%, 10% and 40% of all data and thus 6-fold validation is considered. For cases without validation, training counts 60% while test sets count 40% and the parameters are chosen as $10^4$ for λ in RR and 10 for λ in Lasso. The data are divided randomly to the sets.

The results for Linear models are summarized in Table 2. Exogenous inputs ($6^{th}$ to $10^{th}$ cases) improves the prediction accuracy, especially for test sets. With exogenous inputs introduces, the small MSE $r_{tr,i}$ versus large MSE $r_{te,i}$ ($6^{th}$ case) indicates overfitting. This is because ARX generally has more weights, which means more degree of freedoms. But OLS does not ensure to prevent overfitting. It can be noted that Lasso provides a prediction worse than the baseline case for this problem. For free run prediction, MSE$r_{m,i}$ tends to ∞ which indicates an unstable system. Considering both accuracy and time cost, the $8^{th}$ case, ARX by RR with validation is the best.

**Table 2: Evaluation for Linear Auto-Regressive model**

| Method | $E_i[MSE_{a,i}]$ | $E_i[\text{MSE}r_{te,i}]$ | $E_i[\text{MSE}r_{tr,i}]$ | $E_i[\text{MSE}r_{m,i}]$ | Time (sec) |
|---|---|---|---|---|---|
| AR OLS | 0.0665 | 0.0609 | 0.0742 | 1 | 0.0031 |
| AR RR | 0.0661 | 0.0634 | 0.0707 | 1 | 0.0037 |
| AR RR V | 0.0659 | 0.0628 | 0.069 | 1 | 0.008 |
| AR Lasso | 0.1186 | 0.1197 | 0.1178 | 12.9388 | 0.0042 |
| AR Lasso V | 0.1455 | 0.1429 | 0.1594 | 1 | 0.1373 |
| ARX OLS | 0.0449 | 0.0198 | 0.0842 | 1 | 0.0218 |
| ARX RR | 0.0376 | 0.0329 | 0.0646 | 1 | 0.0207 |
| ARX RR V | 0.037 | 0.031 | 0.0453 | $5:909 \times 10^3$ | 0.0367 |
| ARX Lasso | 0.0908 | 0.0924 | 0.0.0882 | 0.4723 | 0.026 |
| ARX Lasso V | 0.091 | 0.0837 | 0.1001 | 137.7082 | 9.1866 |

**Polynomial Auto-regressive Model**

In this section, we implement polynomial Auto-regressive model training. The setup of maximum time delay, training, validation, and test sets are same as the previous section. For cases without validation, parameters are chosen as $10^4$ for λ in RR and 10 for λ in Lasso. As a trade-off among overfitting, training time, and accuracy, we consider the highest order of polynomial of 2 and without feature interaction terms.

The results for polynomial models are summarized in Table 3. Polynomial models perform generally better than Linear models. According to comparison between 1[th] to 5[th] cases and 6[th] to 10[th] cases, exogenous input provides a better prediction. When exogenous inputs are considered, overfitting occurs for the 6[th] case. These are the same observations in previous section. For free run prediction, MSE$r_{m,i}$ tends towards 1 which indicates an unstable system. Thus, considering both accuracy and time cost, the 8[th] case, ARX by RR with validation is the best.

**Table 3: Evaluation for Polynomial Auto-Regressive model**

| Method | $E_i[MSE_{a,i}]$ | $E_i[\text{MSE}r_{te,i}]$ | $E_i[\text{MSE}r_{tr,i}]$ | $E_i[\text{MSE}r_{m,i}]$ | Time (sec) |
|---|---|---|---|---|---|
| AR OLS | 0.0481 | 0.0432 | 0.0557 | ∞ | 0.0031 |
| AR RR | 0.0495 | 0.0466 | 0.0544 | ∞ | 0.0031 |
| AR RR V | 0.0475 | 0.0454 | 0.0517 | ∞ | 0.0067 |
| AR Lasso | 0.1248 | 0.1227 | 0.1278 | 1.7786 | 0.0104 |
| AR Lasso V | 0.0902 | 0.0865 | 0.0948 | ∞ | 0.6694 |
| ARX OLS | 0.0352 | 0.014 | 0.0655 | ∞ | 0.0175 |
| ARX RR | 0.0298 | 0.0214 | 0.0424 | ∞ | 0.0157 |
| ARX RR V | 0.0326 | 0.0271 | 0.0404 | ∞ | 0.0232 |
| ARX Lasso | 0.0883 | 0.0929 | 0.0824 | 4.2248 | 0.0294 |
| ARX Lasso V | 0.0884 | 0.0845 | 0.0937 | ∞ | 11.4523 |

**Recurrent Neural Network Model**

Here we implemented SRNN and LSTM for time series learning. As previous cases, maximum time delay td of 2 and 4 were considered. For SRNN, we included 10 hidden units for all cases. The networks were trained by Levenberg-Marquardt (LM) optimization. For LSTM, 200 hidden units was considered. Adaptive moment estimation (Adam) algorithm was implemented for network training. The setup of training, validation, and test sets are same as previous section. All algorithm parameters are listed in Table 4. It should be noted that generally there is no guide for parameter determination for both network structures and training algorithms. All parameters selected are obtained by starting from default settings and trying different values.

**Table 4: Training algorithm configurations**

| LM parameters | Values | Adam parameters | Values |
|---|---|---|---|
| Maximum epochs | 1000 | Maximum epochs | 1000 |
| Minimum gradient | $10^{-7}$ | Gradient threshold | 1 |
| Initial μ | 0.001 | Initial learn rate | 0.005 |
| μ decrease factor | 0.1 | Learn rate drop period | 800 |
| μ increase factor | 10 | Learn rate drop factor | 0.2 |
| Maximum μ | $10^{10}$ | Validation frequency | 1 for 50 epochs |

Note: μ is a parameter in LM update formula, μ=0 leads to Newton method while large μ=0 leads to gradient descent with a small step size (MathWorks [2019])

The results for SRNN are summarized in Table 5. Exogenous inputs are extremely important for accurate prediction. In the first two AR cases without exogenous input, the small MSE$r_{tr,i}$ versus huge MSE$r_{te,i}$ indicates overfitting. This can explain why the third and fourth AR case, by introducing validation, behaves much better. However, AR cases generally cannot do free run prediction. With exogenous inputs introduced, the free run prediction for ARX cases (4$^{th}$ to 8$^{th}$ cases) are remarkable. It can be note that adding maximum time delay from 2 to 4 does not significantly increase the prediction accuracy. This is because the error series for cases with maximum time delay of 2 is already almost a white noise (concluded by its autocorrelation). Overfitting still happens in ARX cases without validation (5$^{th}$ and 6$^{th}$ cases) given that MSE$r_{tr,i}$ are much smaller than MSE$r_{te,i}$. Despite the MSE$r_{te,i}$ and MSE$r_{m,i}$ are excellent, the cases are still subject to the risk of overfitting in more complicated cases. Thus, considering both accuracy and time cost, the 7$^{th}$ case, with $t_d$=2 with validation is the best. The corresponding figure is shown in Figure 4.

**Table 5: Evaluation for shallow RNN**

| Method | $E_i[MSE_{a,i}]$ | $E_i[\text{MSE}r_{te,i}]$ | $E_i[\text{MSE}r_{tr,i}]$ | $E_i[\text{MSE}r_{m,i}]$ | Time (sec) |
|---|---|---|---|---|---|
| AR 2 | 5.7678 | 14.2494 | 0.0134 | 191.926 | 3.493 |
| AR 4 | 2.4975 | 6.8467 | 0.0061 | 1688.6 | 4.3046 |
| AR 2 V | 0.0312 | 0.0422 | 0.0245 | 0.405 | 0.1818 |
| AR 4 V | 0.0301 | 0.0407 | 0.0234 | 0.6583 | 0.183 |
| ARX 2 | 0.0147 | 0.0366 | $8.2 \times 10^{-27}$ | 0.0162 | 1.5399 |
| ARX 4 | 0.0135 | 0.0343 | $3.2 \times 10^{-28}$ | 0.0145 | 6.4274 |
| ARX 2 V | 0.0204 | 0.0368 | 0.0091 | 0.025 | 1.4589 |

| ARX 4 V | 0.0214 | 0.0413 | 0.0081 | 0.0232 | 6.3852 |
| AR 2 | 5.7678 | 14.2494 | 0.0134 | 191.926 | 3.493 |
| AR 4 | 2.4975 | 6.8467 | 0.0061 | 1688.6 | 4.3046 |

Note: V is validation, number 2 and 4 indicates td

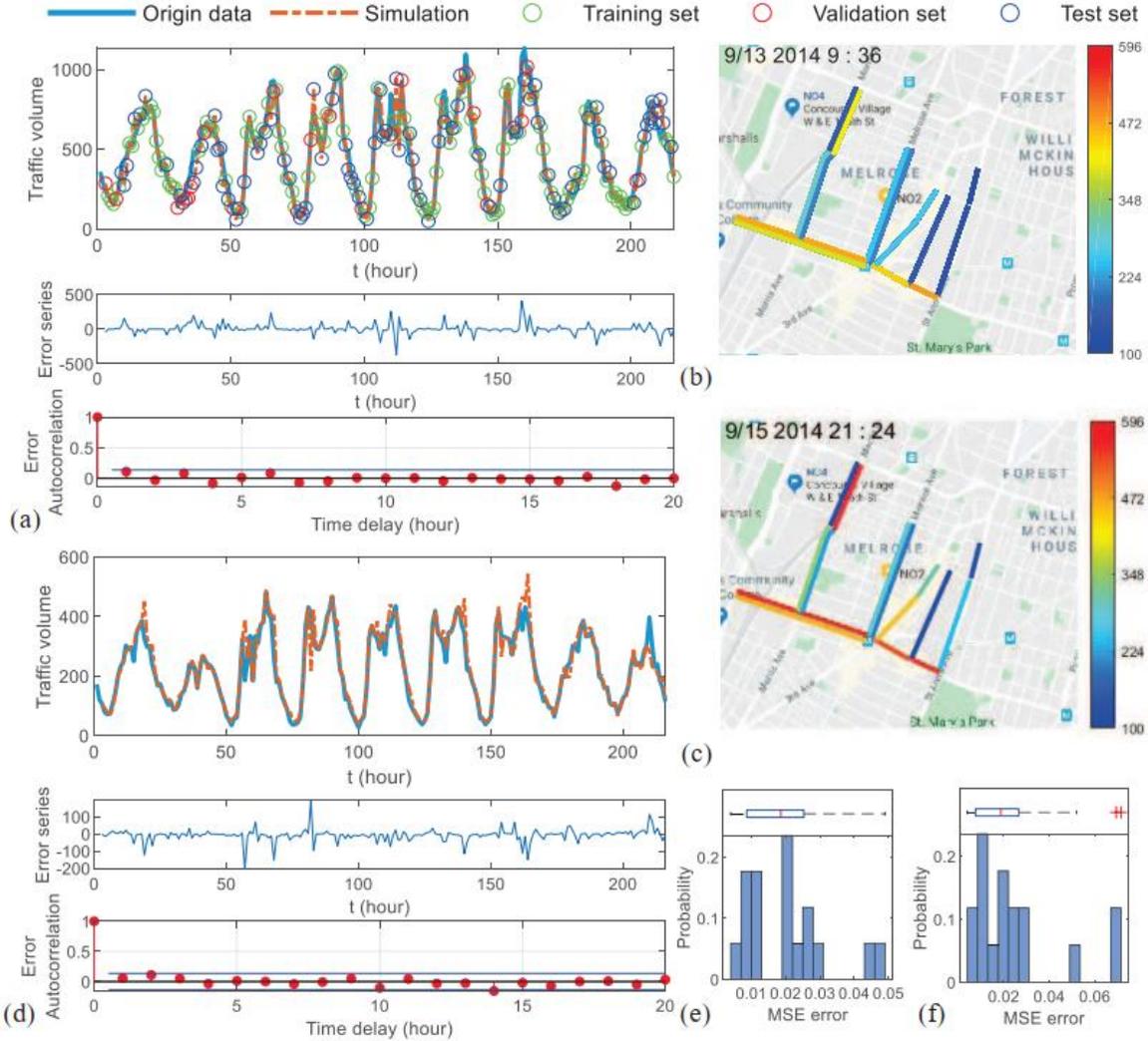

**Figure 4: ARX by SRNN with validation (a) One step ahead prediction evaluation (b-c) One step ahead predicted heat maps at two representative time points (d) Free run prediction (e-f) Statistics of MSEa for one step ahead and free run predictions.**

The results for LSTM are summarized in Table 6. It is impressive to note that no overfitting in the cases, by comparing the $MSEr_{tr,i}$ and $MSEr_{te,i}$. When exogenous inputs are available, SRNN is better than LSTM. This is probably due to that the equation that governing the dynamics of traffic volumes are more like the structure of SRNN than LSTM. Besides, the signal may only have short term dependency. Thus, the SRNN is able to capture accurately the signal dynamic behavior while the LSTM performs worse even with longer time delay (The 5[th] case in Table 6). Considering both accuracy and time cost, the 8[th] case, with $t_d=2$ with validation is the best LSTM. The training time is generally longer for LSTM than SRNN. Thus, for this problem, given exogenous inputs available, SRNN performs better.

**Table 6: Evaluation for shallow LSTM**

| Method | $E_i[MSE_{a,i}]$ | $E_i[\text{MSE}r_{te,i}]$ | $E_i[\text{MSE}r_{tr,i}]$ | $E_i[\text{MSE}r_{m,i}]$ | Time (sec) |
|---|---|---|---|---|---|
| AR 2 | 0.078 | 0.078 | 0.0772 | 0.0761 | 189.32 |
| AR 4 | 0.104 | 0.1029 | 0.1056 | 0.1034 | 111.32 |
| AR 2 V | 0.0747 | 0.0772 | 0.0732 | 0.0739 | 116.76 |
| AR 4 V | 0.0689 | 0.0668 | 0.07 | 0.0679 | 145.67 |
| AR 20 V | 0.07 | 0.0759 | 0.0661 | 0.068 | 255.26 |
| ARX 2 | 0.0299 | 0.0355 | 0.0261 | 0.0295 | 156.65 |
| ARX 4 | 0.0438 | 0.0471 | 0.0413 | 0.0374 | 193.53 |
| ARX 2 V | 0.0351 | 0.0411 | 0.0311 | 0.0345 | 112.24 |
| ARX 4 V | 0.051 | 0.0606 | 0.0446 | 0.0471 | 153.53 |
| AR 2 | 0.078 | 0.078 | 0.0772 | 0.0761 | 189.32 |

Note: V is validation, number 2 and 4 indicates td

## CONCLUSIONS

In this research, we investigated the application of auto-regressive model with exogenous input (ARX) model in real New York city traffic volume prediction. The traffic volume data was firstly presented and analyzed by DTW based K-Means clustering. Linear and polynomial AR or ARX model were training by firstly applying LARs algorithm for feature selection, and then by ordinary least square (OLS), ridge regression (RR) and Lasso (with/without cross validation). Shallow recurrent neural network (SRNN) and long short-term memory (LSTM) were trained by Levenberg-Marquardt (LM) optimization and adaptive moment estimation (Adam) algorithm, respectively. Single step ahead and free run predictions were performed based on the models obtained. Conclusions as follows can be drawn: ARX generally performs better than AR; DTW based K-Means clustering is able to divide data into several clusters with clear signal features; Shallow recurrent neural network ARX is the best trade-off among overfitting, training time, and accuracy; Long short-term memory AR can generate acceptable one step ahead and free run predictions when exogenous inputs are not available.

José Maria P Menezes Jr and Guilherme A Barreto. *Long-term time series prediction with the narx network: An empirical evaluation.* Neurocomputing, 71(16-18):3335–3343, 2008.

New York Metropolitan Transportation Council NYMTC. New York best practice model (npbpm). URL https://www.nymtc.org/Data-and-Modeling/

New-York-Best-Practice-Model-NYBPM. Accessed October 24, 2019.

Billy M Williams. *Multivariate vehicular traffic flow prediction: evaluation of arimax modeling.* Transportation Research Record, 1776(1):194–200, 2001.

N. Zhang, F. Wang, F. Zhu, D. Zhao, and S. Tang. Dynacas. *Computational experiments and decision support for its.* IEEE Intelligent Systems, 23(6):19–23, Nov 2008. doi:10.1109/MIS.2008.101.